\documentclass[pre,reprint,floats,amsmath,amssymb,superscriptaddress,floatfixl,aps]{revtex4-1}

\usepackage[utf8]{inputenc}
\usepackage[T1]{fontenc} 
\usepackage{mathptmx}
\usepackage{fixmath}
\usepackage{amsmath,amssymb}

\usepackage{graphicx}
\usepackage{bm}
\usepackage{color}
\usepackage{hyperref}
\usepackage{ulem}
\usepackage{amsmath}
\usepackage{amssymb}
\usepackage{mathtools}
\usepackage{mathrsfs}

\usepackage{physics}

\usepackage[caption=false]{subfig} % instead of subcaption package

\usepackage[dvipsnames]{xcolor}
\usepackage{tikz}
\usepackage{kotex}

\extrafloats{100}

\begin{document}

\title{Active diffusion of self-propelled particles in flexible polymer networks}

\author{Yeongjin Kim}
%\email{author1@email.com}
\affiliation{Department of Physics, Pohang University of Science and Technology (POSTECH), Pohang 37673, Republic of Korea}

\author{Sungmin Joo}
\affiliation{Department of Physics, Pohang University of Science and Technology (POSTECH), Pohang 37673, Republic of Korea}

\author{Won Kyu Kim}
\email{wonkyukim@kias.re.kr}
\affiliation{School of Computational Sciences, Korea Institute for Advanced Study (KIAS), Seoul 02455, Republic of Korea}

\author{Jae-Hyung Jeon}
\email{jeonjh@postech.ac.kr}
\affiliation{Department of Physics, Pohang University of Science and Technology (POSTECH), Pohang 37673, Republic of Korea}
\affiliation{Asia Pacific Center for Theoretical Physics, Pohang 37673, Republic of Korea}
%\date{\today}

\begin{abstract}
Biopolymer networks having a meshwork topology, e.g., extracellular matrix and mucus gels, are ubiquitous. It is an open question to understand how self-propelled agents such as Janus colloidal particles diffuse through such a biopolymer network. Here, we computationally explore this issue in-depth by explicitly modeling three-dimensional biopolymer networks and performing Langevin dynamics simulations of active diffusion of the self-propelled tracers therein. 
We show that the active tracer performs distinct diffusion dynamics depending on the mesh-to-particle size and P\'eclet number (Pe). When the particle is smaller than the mesh size, it moves as if in free space with a decreased mobility depending on the polymer occupation density and Pe. However, when the particle size is increased to be comparable to the mesh size, the active particles explore the polymer network using the trapped-and-hopping mechanism. We study the trapped time distribution, flight length distribution, the mean-squared displacement, and the long-time diffusivity at varying Pe. If the particle is larger than the mesh, it captures the collective viscoelastic dynamics from the polymer network at short times and the simple diffusion of the total system at large times. Finally, we discuss the scaling behavior of the long-time diffusivity with Pe, where we find a range of Pe that yields a nontrivial power law. The latter turns out to arise from a large fluctuation of trapped, activated tracers in conjugation with responsive polymer networks.
%\cmtWK{I think this finding is important and worth emphasizing, which is not found from ABPs in polymer solutions. We can even more emphasize that this work is the first one studying this.}
\end{abstract}

%\pacs{}

\maketitle

\section{Introduction}

Diffusion of particles exploring through a polymer network is an important subject extensively investigated with a wide range of examples and various motivations. Including the macromolecular diffusion in the chromosome-filled nucleus~\cite{ellenberg2009,vargas2005PNAS,shaban2020HiD}, there are numerous biological examples related to this subject~\cite{Metzler2012PhyTod,franosch2013review,metzler2014anomalous,etoc2018NatMat}. Examples include the transport of lipid granules or purinosomes in the cytoskeleton and endoplasmic reticulum network~\cite{jeon2011PRL,burov2013pnas,french2016purinosome}, extracellular vesicles in extracellular matrix~\cite{lenzini2020matrix,engin2017ECMreview}. 
Diffusion in such a complex network results in intriguing transport phenomena, in which fine-tuned selective filtration can be achieved depending on particle--network interactions and network properties~\cite{stylianopoulos2010diffusion,lieleg2009selective,lieleg2011biological,arends2013ion,lai2010nanoparticles,witten2017particle}.
Probing the particle diffusion is also an essential component in micro-rheology in order to obtain microscopic information of the structure and dynamics of an embedding viscoelastic environment~\cite{MacKintosh1997macromolecules,wong2004PRL}. Moreover, it has important applications in engineering, such as control of molecular permeability in polymer matrix~\cite{vagias2014hydrogels,milster2021tuning,kim2019tuning,kim2017cosolute} and polymer-involved drug deliveries~\cite{amsden1998hydrogels,lu2021NP}   

Currently, an ambitious yet poorly explored subject is the active diffusion of self-propelled particles in such a polymeric complex environment. Here, the self-propelled particles refer to biological or artificial agents that move in a viscous environment with the aid of athermal energy sources by up-taking them from the environment or consuming its own internal chemical energy~\cite{bechinger2016active}, Prominent examples include microswimmers such as E Coli~\cite{bergEColi,Dobnikar2009}, Janus colloidal particles~\cite{bocquet2010Janus}, and molecular motor--macromolecule complexes~\cite{weihs2010,granick2015NM,jeon2018RNAP}. These particles perform persistent random walks with an activity-dependent memory time, which results in Fickian yet active diffusion that violates the Einstein relation~\cite{chen2007prl,maggi2014prl}. The diffusion characteristics of active particles in a viscous fluid have been extensively investigated including the establishment of theoretical modeling~\cite{wu2000PRL,palacci2010prl,leptos2009prl,bricard2013nature,loewen2013PRE}.
Beyond these work, recently, a few papers reported about the diffusion of active tracers in melted polymer solutions~\cite{rajarshi2017Janustracer,yuan2019PCCP,du2019softmatter}. In these studies, the diffusion of active tracers was examined upon the variation of the control parameters, e.g., polymer crowding density, polymer chain length, tracer size, and active force. The coupling of translational and rotational diffusion of the active particles and the nonlinear dependence of tracer's size on the active and drag forces lead to various diffusion dynamics depending on the parameter.  

\begin{figure*}[t]
\includegraphics[width=0.9\textwidth]{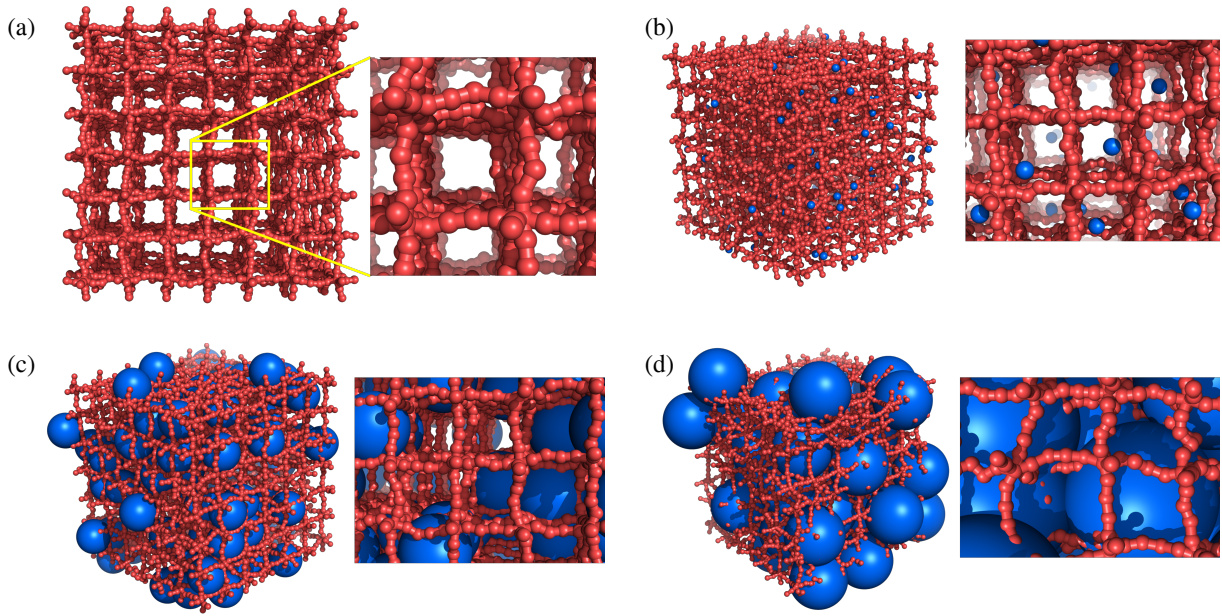}
\caption{\label{fig1}  Illustration of active tracer particles (blue) embedded in a swollen polymer network (red). (a) The constructed cubic polymer network based on a flexible chain model with beads of diameter $\sigma_0$. The cubic cell has dimension of $\ell_0\times\ell_0\times\ell_0$ with $\ell_0=5\sigma_0$. 
%In this figure, the size of the tracer $\sigma_{\mathrm{tr}}$ is comparable to the mesh size $\ell_0 = 5\sigma_0$ where $\sigma_0$ is the polymer bead size. \cmt{NEED a better figure}
In panels (b)--(d), 100 non-interacting AOUP tracers are embedded in this polymer network. 
(b) The case of small AOUP tracers ($\sigma_{\mathrm{tr}}=\sigma_0=0.2\ell_0$, Sec.~\ref{sec3}).  
(c) The case of mesh-sized AOUP tracers ($\sigma_{\mathrm{tr}}=5\sigma_0=\ell_0$, Sec.~\ref{sec4}).
(d) The case of large AOUP tracers ($\sigma_{\mathrm{tr}}=10\sigma_0=2\ell_0$, Sec.~\ref{sec5}).
}
\end{figure*}

In this work, we are interested in the active transport of self-propelled particles exploring a flexible regular polymer network (Fig.~\ref{fig1}). Our polymer system is distinguished from the above melted polymer solution in that the chain is all connected to form a three-dimensional meshwork with a well-defined network topology and mesh size. 
Here the polymer network acts as the fluctuating periodic obstacle or cavity rather than a viscoelastic fluid. Our aim is at understanding the active transport dynamics depending on the mesh-to-particle size ratio by varying the tracer's size. As shown in the studies using the Brownian tracer~\cite{milster2021tuning,kim2022fractal,JSKim2020tightlymeshed}, the accessible volume and its connectivity network drastically differ from the geometrical ratio, which is expected to result in the size-dependent diffusion dynamics for the active tracer. Our model system serves as a prototype model for the study of active diffusion in the abovementioned biological meshwork or artificial regular polymer matrices~\cite{lu2021NPnetwork,cao2021activeNPnetwork}. 
In a broader interest, additionally, our work is a relevant example of the active particle in a periodic confining potential \cite{ribeiro2020abpperiodic} and is intimately related to the study of active particles in a porous matrix~\cite{schwartz2021PNAS}.

The paper is organized in the following. In Sec.~\ref{sec2} we start with explaining the model system investigated in this work. First, we describe how to computationally construct the three-dimensional polymer network shown in Fig.~\ref{fig1}. Then we introduce the active self-propelled tracer exploring the polymer network along with a governing Langevin equation and intrinsic transport properties of this particle. Lastly we explain the simulation protocol employed in our simulation study for the diffusion of active tracers in this polymer network. In the following three sections, we separately present the simulation results with the criteria that the particle size is (i) sufficiently smaller than the mesh size $\ell_0$ (Sec.~\ref{sec3}), (ii) comparable with $\ell_0$ (Sec.~\ref{sec4}), and (iii) much larger than $\ell_0$ (Sec.~\ref{sec5}). It turns out that the transport dynamics is nontrivial in the case (ii), which is our main interest in this work. Finally, in Sec.~\ref{sec6} we discuss some of the main results and summarize the work.

%%%%%%%%%%% Fig 1 %%%%%%%%%%%%%%%%%%%%%
%%%%%%%%%%%%%%%%%%%%%%%%%%%%%%%%%%%%%%%%%%%%

\section{Model}\label{sec2}
\subsection{Polymer network and interactions}

We consider a three-dimensional system of self-propelled tracers in a swollen polymer network (gel) as illustrated in Fig.~\ref{fig1}a. 
The network topology is based on a simple cubic lattice of crosslinkers~\cite{netz1997computer,aydt2000swelling,erbas2015energy,kim2017cosolute} 
between which polymers consisting of 4 monomer beads of identical size $\sigma_{0}$ are crosslinked.
In this swollen cubic gel system, the average mesh size is $\approx 5\sigma_{0}$ (see Fig.~\ref{fig1}a).
%\cmt{YJ: please add the reference in blue and cite them.}\cmtWK{done.}

All neighboring polymer monomers and crosslinkers of distance $r$ are bonded via the bead-spring model with a potential $U_{\mathrm{bond}} = k(r - \ell) ^2$ where $k = 100~k_B T/ \sigma_0^2$ is the spring constant and $\ell = \sigma_0$ is the bond length.

For non-bonded pairwise interactions between particles $i$ and $j$, we use the Lennard-Jones (LJ) potential 
\begin{align}
U_{\mathrm{LJ}}^{ij}(r_{ij})  = 
\begin{cases}
4\epsilon_{ij}
 \qty[
\qty( \frac{\sigma_{ij}}{r_{ij}})^{12} - \qty( \frac{\sigma_{ij}}{r_{ij}})^6 
] - U_c & ,~r_{ij} \le r_c \\
0 & ,~r_{ij} > r_c, 
\end{cases}
\label{eqn_LJ}
\end{align}
where $\epsilon_{ij}$ is the potential depth, $\sigma_{ij} = (\sigma_{ii} + \sigma_{jj})/2$ is the distance parameter, $r_c=2.5\sigma_{ij}$ is the cutoff distance, and the energy shift $U_c =4 \epsilon_{ij} \qty[ \qty(\frac{\sigma_{ij}}{r_c})^{12} - \qty(\frac{\sigma_{ij}}{r_c})^6 ]$ fulfills $U_{\mathrm{LJ}}^{ij}(r_c)=0$. 
The LJ potential imposes an excluded volume to the particles in consideration: 
For polymer monomers `m', crosslinkers `c', and tracers `tr', we use $\epsilon_\text{m,m} = \epsilon_\text{m,c} = \epsilon_\text{c,c} = \epsilon_\text{m,tr} = \epsilon_\text{c,tr}=0.1~k_B T$, which is essentially repulsive~\cite{kim2019tuning}, whereas we use $\epsilon_\text{tr,tr} = 0$.
The latter makes the self-propelled tracers non-interacting among them for simplicity.

\subsection{Active tracers}

We model the translational movement of self-propelled particles with the so-called active Ornstein-Uhlenbeck particle (AOUP). The diffusion dynamics of AOUPs is governed by the following underdamped Langevin equation  \cite{nguyen2021active,caprini2021generalized,shankar2018hidden}
\begin{align}\label{eq:langevin_tr}
    m\dv{\mathbf{v}}{t} & = - \gamma \mathbf{v} - \nabla U_\text{LJ} + \boldsymbol{\xi} + \mathbf{F}_{\mathrm{A}}.
\end{align}
Here, $\gamma$ is the friction coefficient, -$\nabla U_\text{LJ}$ is the LJ force acting on the tracer, and $\boldsymbol{\xi}=(\xi^x,\xi^y,\xi^z)$ is the thermal white noise that fulfills $\langle \boldsymbol{\xi} \rangle=0$ and $\langle \xi^\alpha(t) \xi^\beta(t') \rangle = 2 \gamma k_B T \delta_{\alpha \beta} \delta(t-t')$.
$\mathbf{F}_{\mathrm{A}}=({F}_{\mathrm{A}}^x,{F}_{\mathrm{A}}^y,{F}_{\mathrm{A}}^z)$ is the active Ornstein--Uhlenbeck force with magnitude $\gamma v_p$ ($v_p$ is called the propulsion speed). It is a correlated Gaussian noise of zero mean $\langle \mathbf{F}_{\mathrm{A}}\rangle=0$ and covariance \cite{Joo2020ABP,maggi2015multidimensional,eisenstecken2016conformational}
\begin{align}
\left \langle {F}_{\mathrm{A}}^\alpha (t) {F}_{\mathrm{A}}^\beta (t') \right \rangle &= \frac{\gamma^2 v_p ^2}{3} \delta_{\alpha \beta} e^{-|t-t'|/\tau_{\mathrm{A}}}.
\end{align}
The decay time $\tau_{\mathrm{A}}$ signifies the characteristic time for a persistent motion due to the active force. A trajectory of an active tracer in the overdamped regime has the persistent length $v_p \tau_{\mathrm{A}}$  \cite{bechinger2016active,lowen2020inertial}.
For a free AOUP ($U_\text{LJ}=0$), the mean-squared displacement (MSD) has the analytical form \cite{nguyen2021active}
\begin{align}
&
\begin{aligned}
& \left \langle \Delta \mathbf{r}^2 (t) \right \rangle_\text{free} =  6 D_{\mathrm{th}}\qty(t - \tau_0 \qty(1-e^{-{t}/{\tau_0}})) \\
%+ \frac{2 v_p^{2} \gamma^{2} \tau^{2} \tau_{A}}{m^{2}\left(\tau^{2}-\tau_{A}^{2}\right)} \times \\\\
& + \quad 
2v_p ^2 \tau_{\mathrm{A}} \qty[  \frac{t - \tau_{\mathrm{A}} \qty( 1 - e^{-t/\tau_{\mathrm{A}}})  }{1-\tau_0 ^2 /\tau_{\mathrm{A}} ^2} 
- \frac{\tau_0 ^2}{\tau_{\mathrm{A}} ^2} \frac{t - \tau_0 \qty( 1 - e^{-t/\tau_0})  }{ 1 - \tau_0 ^2/\tau_{\mathrm{A}} ^2 }]
%\left[\left(-1+e^{-\frac{t}{\tau}}\right) \tau^{3}+\left(1-e^{-\frac{t}{\tau_{A}}}\right) \tau_{A}^{3}+t\left(\tau^{2}-\tau_A^{2}\right)\right]
\label{msdfree}
\end{aligned}\\
%& \begin{aligned}
%= & 6 D_{\mathrm{th}}\qty(t - \tau\qty(1-e^{-{t}/{\tau}}))  \\
%& + 2v_p ^2 \tau_{\mathrm{A}} \qty[  \frac{t - \tau_{\mathrm{A}} \qty( 1 - e^{-t/\tau_{\mathrm{A}}})  }{1-\tau^2 /\tau_{\mathrm{A}} ^2} 
%- \frac{\tau ^2}{\tau_{\mathrm{A}} ^2} \frac{t - \tau \qty( 1 - e^{-t/\tau})  }{ 1 - \tau^2/\tau_{\mathrm{A}} ^2 }]
%\end{aligned}\\
& \begin{aligned}
\simeq ~& 6 D_{\mathrm{th}}\qty(t - \tau_0 \qty(1-e^{-{t}/{\tau_0}})) + 2v_p ^2 \tau_{\mathrm{A}} \qty[ 
{t - \tau_{\mathrm{A}} \qty( 1 - e^{-t/\tau_{\mathrm{A}}})  } ],
\end{aligned}\label{msdfree2}
\end{align}
where the last expression is obtained if $\tau_{\mathrm{A}} \gg \tau _0$. In R.H.S., the first term explains the contribution from thermal energy where $D_{\mathrm{th}} = {k_B T}/{\gamma}$ is the thermal diffusivity and $\tau_0 = {m}/{\gamma}$ the momentum relaxation time after which the system becomes overdamped. The second term features the additional effect on the MSD arising from the active force. In this paper, we use $\tau_{\mathrm{A}}=10$ and $\tau_0 = 5/\sigma$ (where $\sigma=1$ for polymer monomers and small AOUPs, $\sigma=4$--$6$ for mesh-sized AOUPs, and $\sigma=10$ for large AOUPs), therefore a persistent time due to the active force is sufficiently longer than the relaxation time.

It is often convenient to use the P\'eclet number to quantify the active propulsion force, which is the ratio of advective transport rate to diffusive transport rate of the mass transport \cite{bechinger2016active},
\begin{align}
\operatorname{Pe}= \frac{\sigma_\text{tr} v_p}{D_{\mathrm{th}}} = \frac{3\pi \eta \sigma_{\mathrm{tr}}^2 v_p }{k_B T},
\end{align}
where $\gamma = 3 \pi \eta \sigma_\text{tr}$ ($\eta$: viscosity) is used in the last expression. In our study, Pe is used for the measure of the active strength as a key parameter.

\subsection{Simulation and parameters}

We run Langevin dynamics simulations in an $NVT$ ensemble by employing the LAMMPS package~\cite{plimpton1995fast}. For the $i$-th network particle (monomers and crosslinkers) we employ the normal Langevin equation,
\begin{align}\label{eq:langevin_ptr}
    m\dv{\mathbf{v}_i}{t} & = - \gamma \mathbf{v}_i - \nabla_i U + \boldsymbol{\xi}_i,
\end{align}
where $-\nabla_i U$ is the force due to the total potential $U=\sum_{i,j} [U_\text{bond}+U_\text{LJ}]$ acting on the $i$-th particle.
For the AOUP dynamics Eq.~\eqref{eq:langevin_tr} is implemented by modifying the package script.

We use the LJ units, thereby having the unit length $\sigma_0$ and the unit time $t_0=\sqrt{m \sigma_0^2/k_B T}$, which are set equal to unity. A periodic simulation box of size $35\times 35\times 35$ is considered where $N=100$ active tracer particles of size $\sigma_\text{tr} \equiv \sigma_\text{tr,tr}$ are immersed in a swollen polymer network, i.e., 7 crosslinkers and 7 polymers (of 4 monomers) per line (see Fig.~1a--d).
We use the time step $\delta t =0.001$ and $T=5\times 10^7$ total time steps.
%So the maximum time that we simulated is $50000$.
%Since the hopping is rare event for passive tracers, simulating $T=5\times 10^4$ does not give sufficient statistics. So we run additional simulations for passive tracers of size $\sigma_{\mathrm{tr}}=5$ to estimate the mean sojourn time more accurately.\cmtWK{more accurately? or as reference ?}

%In this model, we compared the active tracer size $\sigma_{\mathrm{tr}}$ and mesh size of the polymer network $\ell_0$. 
We focus on the diffusion dynamics of the active tracers depending on the tracer size and the self-propulsion activeness. To this end, we consider size parameters for small tracers $\sigma_{\mathrm{tr}} \ll \ell_0$, mesh-sized tracers $\sigma_{\mathrm{tr}} \sim \ell_0$, and large tracers $\sigma_{\mathrm{tr}} \gg \ell_0$ in the polymer network of mean mesh size $\ell_0 = 5$. We adopt $\sigma_{\mathrm{tr}} = 1$ for the first case, $\sigma_{\mathrm{tr}} = 4$--$6$ for the second case, and $\sigma_{\mathrm{tr}} = 10$ for the last case. %For the intermediate-sized tracers, we use $\sigma_{\mathrm{tr}}=3.0,3.5,4.0,4.5,5.0,5.5$ and $6$.
For each tracer size, we vary the P\'eclet number as another key parameters.  
%\cmt{Provide Pe number.} \cmtWK{Maybe better to present a table for all parameters here...}
For the small tracer, we use $\operatorname{Pe}=0$--$10$. For the intermediate tracers, we use $\operatorname{Pe}=0$--$180$. For the large tracer, we use $\operatorname{Pe}=0$--$600$.

%%%%%%%%%%%%%%%%% Fig 2 %%%%%%%%%%%%%%%%%
\begin{figure*}
\includegraphics[width=0.7\textwidth]{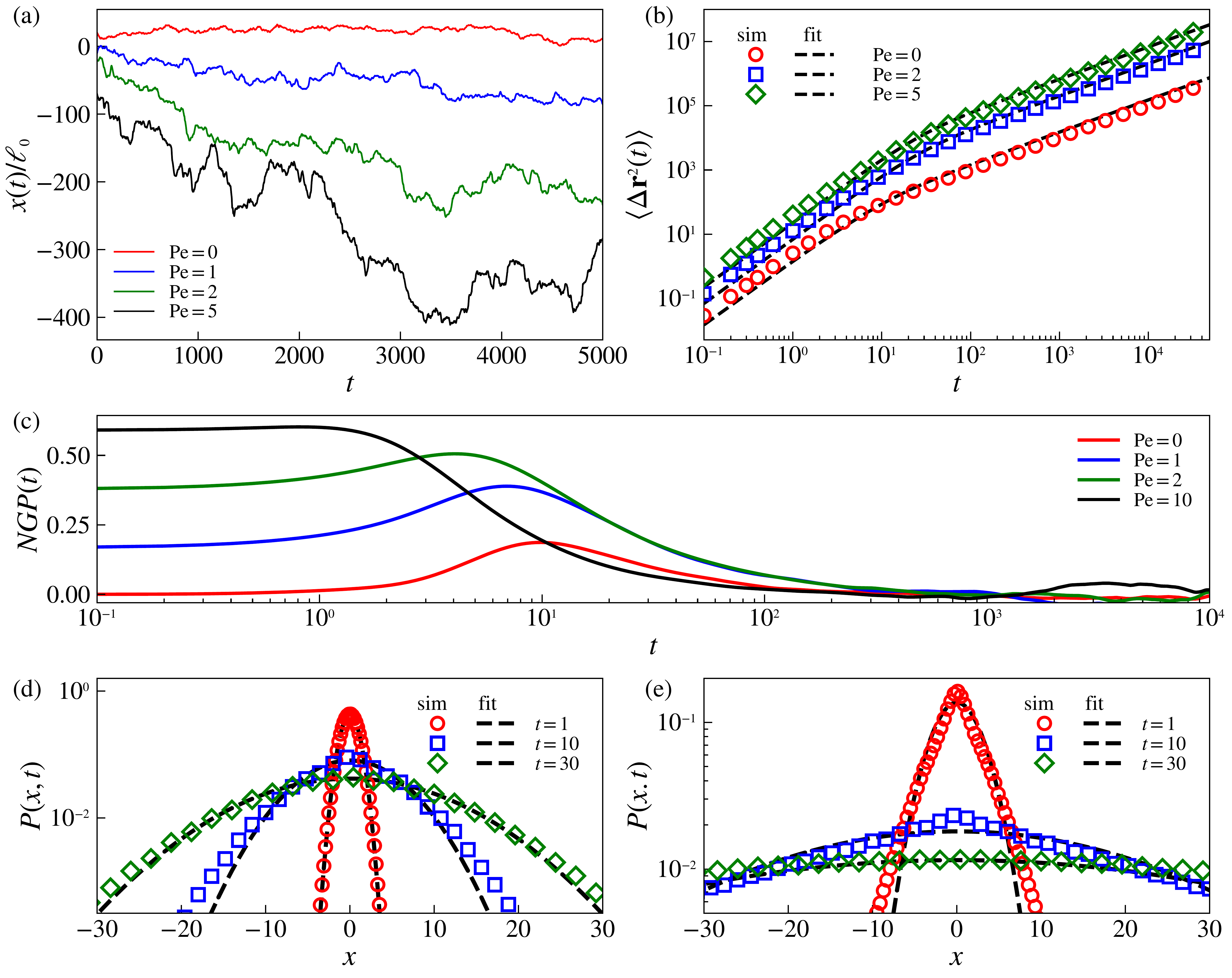}
\caption{\label{fig2} Active diffusion for a small particle (of diameter $\sigma_{\mathrm{tr}}=1$) embedded in the polymer network (of mean mesh size $\ell_0=5$). (a) Sample single-particle trajectories (showing $x$-component diffusion). 
(b) The MSDs from the simulation at various Pe numbers. The dashed lines denote the corresponding fit curves with $C\times$ MSD in Eq.~\eqref{msdfree} where $C$ is a fit parameter for each Pe number. 
%In the simulation the MSD is calculated via the formula $\langle \Delta \mathbf{r}^2(t)\rangle=\frac{1}{N}\sum_{i=1}^N \frac{1}{T-t}\int_{0}^{T-t} d\tau \left( \mathbf{r}_i(t+\tau) - \mathbf{r}_i(\tau) \right)^2$ where $N=100$ and $T=5\times 10^4 t_0$.
(c) The variation of non-gaussian parameter (NGP), $\langle \Delta x^4(t)\rangle/3\langle \Delta x^2(t)\rangle^2-1$, as a function of time lag $t$. 
(d) The van-Hove correlation functions $P(x,t)$ for passive Brownian tracers ($\operatorname{Pe}=0$). %Numerically, it is defined as $P(x,t) = \frac{\mathcal{C}^{-1}}{T-t} \int_0 ^{T-t}d\tau \sum_{i=1}^{N} \delta\qty([ x_i(\tau+t) -x_i(\tau)] - x)$ where $\mathcal{C}$ is the normalization factor. 
%The dashed line shows the gaussian fit to the data. 
(e) The van-Hove correlation functions $P(x,t)$ for active tracers ($\operatorname{Pe}=5$).
The dashed lines in (d) and (e) denote the gaussian fit for each $t$.
}
\end{figure*}
%%%%%%%%%%%%%%%%%%%%%%%%%%%%%%%%%%%%%%%%

\subsection{Analysis}

Here we introduce the physical observables that are extensively investigated in the analysis of the simulation data. To characterize the dynamics of tracers, the mean-squared displacement (MSD) is calculated via the time- and ensemble-averaged formula $\langle \Delta \mathbf{r}^2(t)\rangle=\frac{1}{N}\sum_{i=1}^N \frac{1}{T-t}\int_{0}^{T-t} dt' \left( \mathbf{r}_i(t+t') - \mathbf{r}_i(t') \right)^2$.

To examine the non-gaussianity of motion of tracers, we compute the non-gaussian parameter (1D) $\mathrm{NGP}=\langle \Delta x^4(t)\rangle/3\langle \Delta x^2(t)\rangle^2-1$. The NGP is zero for gaussian motion and has positive values if the tail is thicker than the gaussian. Additionally, we measure the van-Hove self-correlation function $P(x,t)=\frac{\mathcal{C}^{-1}}{T-t} \int_0 ^{T-t}dt' \sum_{i=1}^{N} \delta\qty([ x_i(t'+t) -x_i(t')] - x)$ where $\mathcal{C}$ is the normalization factor. 

We also study physical observables, such as the trapped time distribution $P(\tau)$, flight-length distribution $P_\mathrm{fl}(l)$, and the long-time diffusivity $D_L= \lim_{t\to \infty} {\left \langle \Delta \mathbf{r}^2 (t) \right \rangle}/{(6t)}$, which are defined and discussed in the following sections.  

\section{Active diffusion for small particles}\label{sec3}

We start with the case that the active tracer  is sufficiently smaller than the average mesh size $\ell_0$ (Fig.~\ref{fig1}b). Figure~\ref{fig2}a shows the simulated ($x$-component) trajectories of AOUP tracers ($\sigma_\mathrm{tr}=1$) exploring the polymer network ($\ell_0=5$) at several Pe values. We plot the corresponding MSD curves in the panel (b). A general trend is that the active tracers move in a fashion very similar to the free-space motion at lengthscales of $|\Delta \mathbf{r}|\gtrsim \ell_0$, in which the polymer network acts as a trivial obstacle. For a given cross-over timescale $t^*$, the AOUPs exhibit ballistic movement for $t<t^*$ and Fickian diffusion for $t>t^*$. Here $t^*$ is comparable to $\tau$ for small Pe and $\tau_A$ for large Pe. Qualitatively, the AOUP motion in this case is akin to the free-space case described by the MSD [Eq.~\eqref{msdfree} or \eqref{msdfree2}]. This allows us to compare the simulation data with Eq.~\eqref{msdfree} using a free fitting prefactor $C$, assuming that that the long-time diffusivity [$D_L=\lim_{t \to \infty}\langle \Delta \mathbf{r}^2(t)\rangle_\text{free
}/(6 t)$]
\begin{equation}\label{eq8}
D_L=D_{\mathrm{th}}\qty(1+\frac{D_{\mathrm{th}}\tau_A}{3\sigma_\mathrm{tr}^2}\mathrm{Pe}^2)
\end{equation}
in free space is simply decreased in amplitude by the factor of $C$ because of the obstacles in space. Indeed, the simulation data for MSD are excellently explained by this simple theory, $C \times \langle \Delta \mathbf{r}^2(t)\rangle_\text{free
}$ using Eq.~\eqref{msdfree}, (dashed lines in Fig.~\ref{fig2}b) in the regime of  $|\Delta \mathbf{r}|> \ell_0$. The fitted prefactor $C(\text{Pe})$ is an intriguing function of Pe [see Fig.~\ref{Pe_C_data} in Appendix for $C(\text{Pe})$], which we will discuss more in detail later regarding the diffusivity. 
In terms of percolation theory, the geometric condition is understood such that the entire accessible volume in the polymer network is connected well to such small tracers~\cite{bunde2005chapter,kim2022fractal}. Here, the tracer has three distinct diffusion dynamics depending on the lengthscale \cite{bunde2005chapter,havlin1987review}: It exhibits the free-space motion when $|\Delta \mathbf{r}|/\ell_0\ll 1$. Then it is accompanied by the obstacle-induced transient anomalous dynamics at $|\Delta \mathbf{r}|/\ell_0\approx 1$. Finally, in the long-time limit the tracer reaches the Fickian dynamics with an obstacle-dependent reduced diffusivity $D_L<D_\mathrm{th}$. Our study suggests that active particles have qualitatively the same transport tendency in a percolated geometry.

In Figs.~\ref{fig2}c--e we evaluate the non-gaussian parameter NGP and the van-Hove autocorrelation functions $P(x,t)$ for the Brownian and active OU tracers. The trapped Brownian particle has the expected non-gaussian behavior in the following \cite{SongUmJeon2019}: At short times where $|\Delta \mathbf{r}|\ll \ell_0$, it rarely feels the obstacles, thus manifesting gaussian dynamics [see $\text{NGP}(t\ll \ell_0^2/D_\text{th})\approx 0$ for $\text{Pe}=0$ in Fig.~\ref{fig2}c, and $P(x,t=1)$ in Fig.~\ref{fig2}d]. Then the NGP increases with time and reaches a maximum  at $|\Delta \mathbf{r}|\approx \ell_0$, which is the length that the obstacle effect is the largest. Beyond this length, the diffusion approaches gaussian again because the diffusive trajectory is more similar to a random walk at larger lengths. 

The non-gaussianity is more pronounced for active tracers. Especially, the NGPs become significantly large compared to the Brownian case at short times, in which the NGP tends to be larger with Pe (see Fig.~\ref{fig2}c). Its origin stems from the increased frequency of collisions with the polymer obstacle when Pe is increased. The collision effect can be seen in Fig.~\ref{fig2}e such that $P(x,t)$ with Pe$=5$ gets a sharper cusp at $x=0$ compared to the Brownian counterpart with Pe$=0$ (Fig.~\ref{fig2}d). The NGP has the maximum state at the time where the tracer's explored length is about $|\Delta \mathbf{r}|\approx\ell_0$. After this point, as the Brownian particle shows, the NGP monotonically decreases with increasing $t$ at $|\Delta \mathbf{r}|>\ell_0$ where the explored space looks more homogeneous and continuous. The NGP converges to zero earlier with larger $\mathrm{Pe}$ in time, indicating that the highly active motion approaches gaussianity faster. 

%We find that the transport dynamics of small active tracers share  

%\begin{figure}
%\includegraphics[height=6cm]{vanHove_sigmatr1.0af0.0_log_stix.pdf}
%\includegraphics[height=6cm]{vanHove_sigmatr1.0af0.0_log_stix_nonGaussian.pdf}
%\caption{\label{vanHove_small_passive}The van Hove autocorrelation function for small passive particle: $\sigma_{\mathrm{tr}}=1$ and $\operatorname{Pe}=0$. The van Hove autocorrelation function is defined as $P(x,\Delta) = \frac{\mathcal{C}^{-1}}{T-\Delta} \int_0 ^{T-\Delta} \sum_{i=1}^{N} \delta\qty([ x_i(t+\Delta) -x_i(t)] - x)$. Here $\mathcal{C}$ is the normalization factor. 
%The van Hove autocorrelation function is fitted to Gaussian distribution. \cmtYJ{plots are from new data}}
%\end{figure}

%\begin{figure}
%\includegraphics[height=6cm]{vanHove_sigmatr1.0af27.0_log_stix.pdf}
%\includegraphics[height=6cm]{vanHove_sigmatr1.0af27.0_log_stix_nonGaussian.pdf}
%\includegraphics[height=6cm]{vanHove_sigmatr1.0af5.0_stix.pdf}
%\includegraphics[height=6cm]{vanHove_sigmatr1.0af5.0_log_stix.pdf}
%\caption{\label{vanHove_small_passive}The van Hove autocorrelation function for small active particle: $\sigma_{\mathrm{tr}}=1$ and $\operatorname{Pe}=5.4$. The van Hove autocorrelation function is defined as $P(x,\Delta) = \frac{\mathcal{C}^{-1}}{T-\Delta} \int_0 ^{T-\Delta} \sum_{i=1}^{N} \delta\qty([ x_i(t+\Delta) -x_i(t)] - x)$. Here $\mathcal{C}$ is the normalization factor. % In figure2_11.pdf the Pe is changed to 5.0 
%The van Hove autocorrelation function is fitted to Gaussian distribution. \cmtYJ{plots are from new data}}
%\end{figure}

\section{Active diffusion for mesh-sized tracers}\label{sec4}

In this section, we investigate the transport dynamics of active tracers whose diameter is comparable to the mesh size of the polymer network ($\sigma_\text{tr} \approx \ell_0$), which becomes drastically different from the previous case ($\sigma_\mathrm{tr}<\ell_0$). 

\subsection{Trapped and hopping diffusion}
\begin{figure*}
\includegraphics[width=\textwidth]{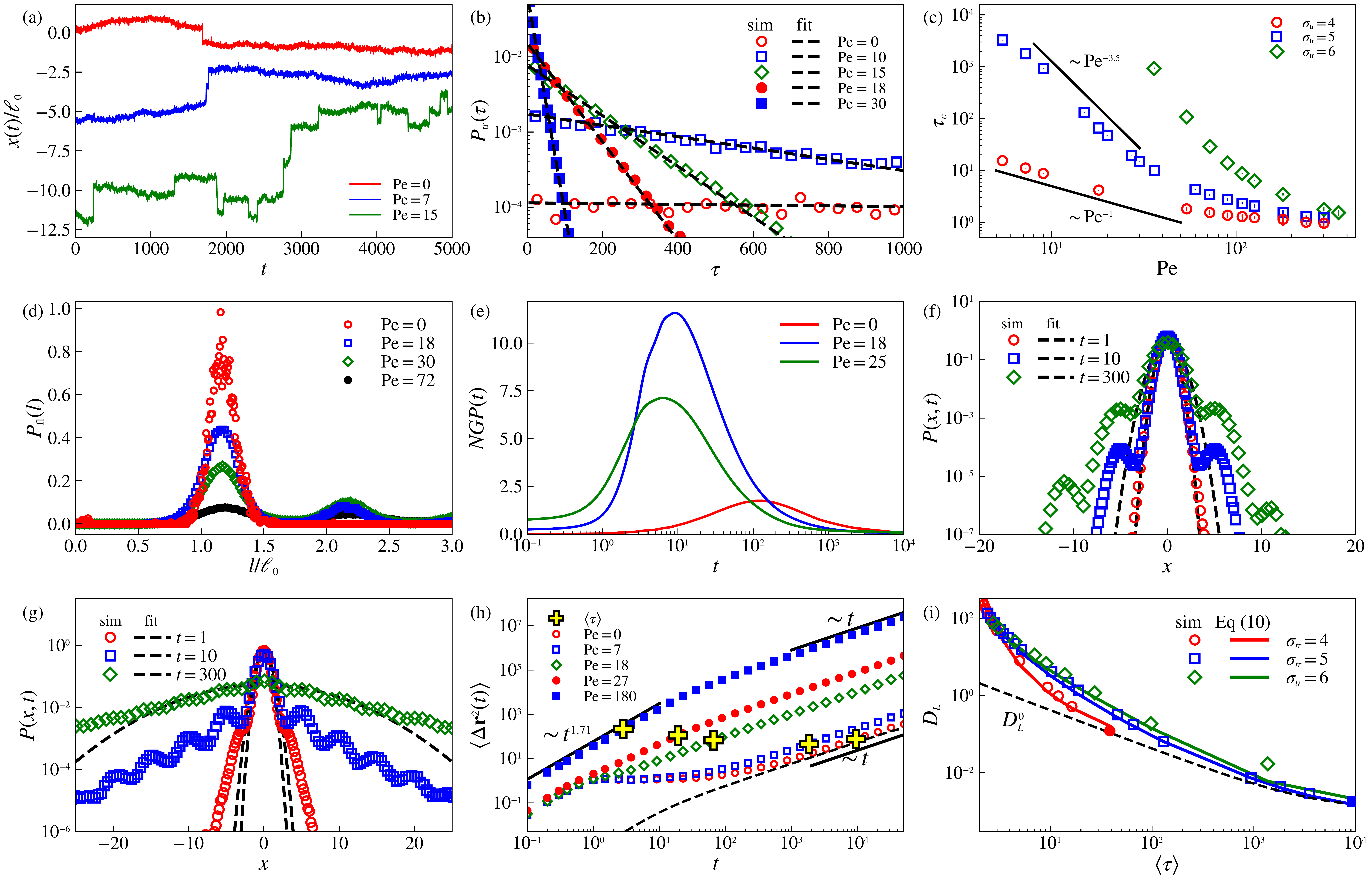}  
\caption{\label{fig3}Active diffusion for mesh-sized particles (of size $\sigma_\mathrm{tr}=4$--$6$) embedded in the polymer network of mesh size $\ell_0=5$. (a) The trapped-and-hopping motion of tracers of  $\sigma_\mathrm{tr}=5$ at Pe$=0,~5$, and 25. 
(b) Trapped time distributions for tracers of $\sigma_\mathrm{tr}=5$. The dashed line shows the exponential distribution Eq.~\eqref{Ptrap} with $\tau_c$ fitted from the data. 
%%%% protocols moved
%From the simulated trajectories we extract the statistics of hopping times and flight lengths by the following way. First, we reduce the roughness in the trajectory using the bilateral filter [cite], making a noise-reduced trajectory $\tilde{x}(t)$ from the original trajectory $x(t)$. We define the state of the particle every unit time $t_0$: If $\qty| \tilde{x}(t+3t_0) -\tilde{x}(t)|>3.0\sigma$, the particle's state in $[t,t+t_0)$ is set to be in the hopping state. Additionally, when $x(t+1) - x(t)$ and $\tilde{x}(t+1) - \tilde{x}(t)$ has the same sign to $\tilde{x}(t) - \tilde{x}(t-1)$, state at $t$ is also hopping. Likewise, when $x(t) - x(t-1)$ and $\tilde{x}(t) -\tilde{x}(t-1)$ has the same sign to $\tilde{x}(t+1) - \tilde{x}(t)$, the state at $t$ is hopping. If we find $t_{i}, t_{i+1} , \cdots t_{j}$ such that $t_{i}$ is not hopping state, but $t_{i+1}, t_{i+2}, \cdots ,t_{j}$ are hopping states and $t_{j+1}$ is not hopping state, we define hopping time $t_{j}-t_{i}$. Also, the flight length is defined as $\tilde{x}(t_{j}) - \tilde{x}(t_{i})$. Likewise, if we find $t_{i}, t_{i+1} , \cdots t_{j}$ such that $t_{i}$ is hopping state, but $t_{i+1}, t_{i+2}, \cdots ,t_{j}$ are not hopping states and $t_{j+1}$ is hopping state, we define trapping time $t_{j} - t_{i}$. 
%%%%%%%%%%%%%%%%%%%%%%%%%%
(c) The fitting parameter for the mean trapped time $\tau_c$ in Eq.~\eqref{Ptrap} vs. Pe for different $\sigma_\text{tr}$.
(d) The flight length distributions of tracers of $\sigma_\mathrm{tr}=5$ at Pe=0, 18, 30, and 72.  
(e) The plot of $\mathrm{NGP}(t)$ of tracers (of $\sigma_\mathrm{tr}=5$) at $\mathrm{Pe}=0$, 18, and 25. 
(f) The van-Hove autocorrelation functions $P(x,t)$ of Brownian particles ($\sigma_\mathrm{tr}=5$) for different lag times $t$.
(g) $P(x,t)$ of AOUPs at $\mathrm{Pe}=18$ for different time lag $t$.
(h) The MSD curves of tracers of $\sigma_\text{tr}=5$ under various Pe conditions. The cross symbol marks the mean trapped time $\langle \tau \rangle$ averaged from the simulation data shown in (b). Three solid lines show the guided scaling relation.
The dashed line depicts the MSD of the center of mass of entire system.
(i) The long-time diffusivity of tracers $D_L$ vs. the average trapped time $\langle \tau \rangle$. The solid and dashed lines represent Eqs.~\eqref{eq:D_L} and \eqref{D0}, respectively.}
\end{figure*}

Figure~\ref{fig3}a shows sample trajectories of active tracers of volume $\sigma_\mathrm{tr}=5(\approx\ell_0)$ at various Pe numbers. In this regime, the tracer size begins to play a role, where tracers are for most times geometrically trapped within a cubic-shaped mesh of size $\ell_0\times\ell_0\times \ell_0$. The trapped particles then occasionally escape from the mesh due to the active propulsion and/or geometrical thermal fluctuations of the polymer mesh, then hopping to neighbor sites. The observed trapped-and-hopping diffusion is reminiscent of the diffusion pattern of Brownian tracers embedded in a polymer network that are reported in experimental or computational studies \cite{wong2004PRL,JSKim2020tightlymeshed,chen2020nanoparticle,Kob2021NPdiffusion}. Because the tracer is tightly trapped in the simulation, the hopping process turns out to be a rare event for the Brownian tracer. 
For the active tracers, however, hopping events are more frequently observed during the simulation, particularly for larger Pe (see the green line Fig.~\ref{fig3}a). The diffusion of the active tracers is kind of tug-of-war between self-propelled propulsion and geometrical trap by the polymer network.  
It is noteworthy that the sample trajectories show that the active tracers can have flights of length larger than the nearest neighbor distance, i.e., $\ell_0$. Such events are more activated as Pe is increased. Meanwhile, the Brownian tracer (Pe$=0$) mostly undergoes the nearest neighbor hopping process.

%%%%%%%%%%%% Fig 3 %%%%%%%%%%%%%%%%%

%\begin{figure}
%\includegraphics[height=6cm]{vanHove_sigmatr5.0af0.0_log_stix.pdf}
%\includegraphics[height=6cm]{vanHove_sigmatr5.0af5.0_log_stix.pdf}
%\caption{\label{vanHovemiddle} The van Hove autocorrelation function for middle monomer case. Left is for passive case, and right plot is for active case with $\operatorname{Pe}=25$.}
%\end{figure}

\subsection{Trapped time distribution and mean trapped time}

We investigate the statistics of trapped times from the simulation trajectories (see the Appendix~\ref{appendixb} for the technical detail about data pre-processing and numerical procedure of extracting the hopping events).

Figure~\ref{fig3}b shows the trapped (i.e., inter-event) time distribution $P_\mathrm{tr}(\tau)$ for AOUPs of $\sigma_\text{tr}=5$ at various Pe values, which is well fitted by an exponential law (dashed line)
\begin{equation}
    P_\mathrm{tr}(\tau)=\tau_c^{-1}e^{-\tau/\tau_c},
    \label{Ptrap}
\end{equation}
where $\tau_c(\mathrm{Pe})$ is the characteristic time. % equal to the mean trapping time $\langle \tau \rangle$. 
This implies that the hopping process is essentially random with the mean trapped time $\tau_c$.
More active the tracer shorter the exponential tail. 
In Fig.~\ref{fig3}c we examine the dependence of fitted $\tau_c$ on Pe for several tracer sizes $\sigma_\text{tr}=4$, 5, and 6. Here, the error bar is smaller than the symbol size.
%The characteristic trapped time overall decays faster than exponentially with the increase of Pe. 
The $\tau_c$ is observed to have two distinct dependence on Pe. Notably, the self-propulsion force dramatically reduces the trapped time for $\mathrm{Pe}\lesssim 50$, particularly for $\sigma_\text{tr}=4$ and 5, after which the hopping process is boosted by the large active force. This implies that there is a threshold for Pe that liberates the trapped particles. 
%It becomes shorter as the tracer is driven by larger distribution is a characteristic of continuous-time Markov chain. This result implies that, in average, the instantaneous probability to jump to other mesh does not depend on the tracer's arrived time $t_{\mathrm{a}}$. 

For additional information, we compare $\tau_c$ to the mean trapped time $\langle \tau\rangle$ obtained from $\int  \tau P_\mathrm{tr}(\tau)d\tau$. See $\tau_c$ vs. $\langle \tau\rangle$  in the Appendix~\ref{appendixc} (Fig.~A3). While both quantities agree with each other for $\langle \tau\rangle\gtrsim 10$, $\tau_c$ becomes greater than $\langle \tau\rangle$ for $\langle \tau\rangle\lesssim 10$. This may imply that for highly activated AOUPs (e.g., for Pe$\gtrsim 50$ and $\sigma_\mathrm{tr}=5$)  $P_\mathrm{tr}(\tau)$ decays faster than the exponential law. Nevertheless, it is found that the decay of $\langle \tau \rangle$ has Pe-dependencies consistent to that of $\tau_c$ (see Fig.~A4 in the Appendix~\ref{appendixd}).    

\subsection{Flight length distribution and multiple-mesh hopping}
We also measure the flight length $(l)$ in the hopping event. Figure~\ref{fig3}d shows the flight length distribution $P_\mathrm{fl}(l)$ for the Brownian particles and AOUPs of size $\sigma_{\mathrm{tr}}=5$. 
The distributions in general have multiple peaks at around the distances of integer times mean mesh size $\langle \ell_0 \rangle$. For the Brownian tracer (Pe$=0$), the flight length is almost restricted to jumps to the nearest meshes. %\cmt{But I can see a small peak at $l<\ell_0$ Why?} \cmtWK{looks like it is the small non-zero probability for BP to stay in the same mesh.} 
The active tracer with its self-propulsion energy can travel multiple mesh distances in a hopping event. Evidently, the propensity of large jumps is increased with Pe. By the same reason, the height of the nearest neighbor flight is decreased as Pe increases.

As shown in Fig.~\ref{fig3}e, the non-gaussian parameter provides further information on the hopping dynamics. The increase of NGP in time indicates that the tracers start to feel the polymer network as obstacles at the corresponding time scale, i.e., the onset of hopping transition. Physically, the time at the peak position of NGP can be understood as the tracer's trapped time. However, we find that this holds only for highly active particles whose trapped time is much shorter than that of Brownian tracers. The main reason is that for these tightly confined (Brownian) tracers, the dynamics of polymer network comes into play particularly for long-time diffusion where the whole network motion can be larger than the rare tracer hopping transition. This polymer network dynamics decreases NGP in the long-time regime apart from the presence of the passive hopping events.        

The tracer hopping events indeed can be seen clearer from the van-Hove distribution.
The Brownian tracers, on one hand, mostly undergo the nearest-neighbor hopping (Fig.~\ref{fig3}f).
On the other hand, AOUPs with a large Pe number ($\mathrm{Pe}=18$) can jump between distant network meshes at time lag $t=10$ (see the blue square symbols in Fig.~\ref{fig3}g) that is around the time of the peak position of NGP. At longer time lag, $t=300$, the oscillatory feature of $P(x,t)$ is suppressed and smoothened by both of the tracer diffusion and the polymer network diffusion. Eventually, in the infinitely-long time regime, the dynamics will be gaussian, which solely originates from the active hopping process with $\text{Pe}>0$.        
\subsection{Mean-squared displacement}   

Figure~\ref{fig3}h shows MSD of the active tracers of size $\sigma_\text{tr}=5$ under various Pe conditions. In the plot the average trapped time $\langle \tau(\mathrm{Pe})\rangle$ is annotated (cross symbols) for reference. It is found that transport dynamics highly depend upon the strength of active propulsion. When the active noise is turned off ($\mathrm{Pe}=0$), the tracer's dynamics essentially represents the trapped motion. After the ballistic regime ($t<1$) it exhibits the confined dynamics within a mesh up to $t\sim 10^2$. Finally it has Fickian motion for $t\gtrsim 10^3$. Note that the main mechanism for the Fickian dynamics is not the Brownian hopping movement of the tracer. Such hopping events are negligible until $t\sim \langle \tau\rangle$ which is the order of $O(10^4)$. As we discussed via NGP and $P(x,t)$ previously, the long-time diffusion is mostly attributed to the drift of the total system, which is confirmed by the MSD of the center of mass (dashed line).  
When the active noise is turned on, we observe three distinct dynamic patterns depending on $\mathrm{Pe}$. When $\mathrm{Pe}\lesssim 18$, the active tracer suffers the transient confinement-induced subdiffusion and cross-overs to the Fickian diffusion after $t\gtrsim\langle \tau \rangle$. We note that at $\mathrm{Pe}\approx 18$ the active particle exhibits a seemingly confinement-free Fickian diffusion. At this special strength, the self-propulsive motion precisely cancels out the geometrical trapping.   
If $\mathrm{Pe}\gtrsim 18$, the active particle features a superdiffusion for $t\lesssim\tau_A=10$ and slows down to a normal diffusion for $t\gtrsim\tau_A$. In this regime, the self-propulsion energy is too high for the tracer to be trapped in a polymer mesh. The fact that the superdiffusion is sub-ballistic ($\alpha\approx1.7$) indicates the strong polymer-induced friction occurs during the propagation.      

It should be noted that the active particle, regardless of the magnitude of Pe, ends up in the Fickian diffusion that stems from the active hopping mechanism. 
The number of hopping events $t/\langle \tau \rangle$ in the Fickian regime gets more significant with Pe.

%The subdiffusive motion of the tracer by confinement of polymer network meshes are the characteristic of intermediate particle and large particle dynamics. For small particles with $\sigma_{\mathrm{tr}} =1$, there's no subdiffusive dynamics since the polymer network meshes cannot trap the small particles. Also, for the large particles with $\sigma_{\mathrm{tr}}=10$, the transport of tracers through network meshes is rare event so we cannot observe the transport in our timescale. The intermediate particles are affected by both events.

\subsection{Long-time diffusivity}
From the viewpoint of random walks, diffusivity is 
$D = \frac{\mathscr{L}^2}{6\mathscr{T}}$
where $\mathscr{L}$ and $\mathscr{T}$ are, respectively, the average jump length and waiting time of the random walker. For our active systems, $\mathscr{L}$ is the average flight length and $\mathscr{T}$ is the sum of average trapped and flight times. Accordingly, we can suggest the long-time diffusivity to be given by
\begin{align}
D_L=\frac{\left \langle \ell^2 \right \rangle}{6 \left[ \langle \tau \rangle + \langle \tau_\mathrm{fl} \rangle \right]} + D_\mathrm{com}.   
\label{eq:D_L}
\end{align}
Here, $D_\mathrm{com}$ refers to the drift of the center-of-mass of the total system, which would be non-vanishing in the laboratory frame. As observed from the simulation, in the limit of $\mathrm{Pe\to0}$ (the Brownian tracer) the tracer's diffusion was found to be dominated by the trapping and hopping into the nearest mesh. Thus, as a special case, we also define the Brownian-limit $D_L$ as
\begin{equation}
D_L^0=\frac{  \ell_0^2  }{6  \langle \tau \rangle}+ D_\mathrm{com},
\label{D0}
\end{equation}
where $D_\mathrm{com}\simeq\frac{1}{N_\mathrm{gel}}\frac{k_BT}{3\pi\eta \sigma_0}$ with $N_\mathrm{gel}$ being the total monomer number of the polymer network. Note that the flight time at $\mathrm{Pe=0}$ is negligible compared to the average trapped time, which results in $D_L^0-D_\mathrm{com}$ having an inverse proportionality relation with the trapping time.

In Fig.~\ref{fig3}i we show the measured relationship between $D_L$ and $\langle \tau \rangle$, which is compared with the above theory. 
It is evident that $D_L$ for the active tracer does not follow the inverse power-law scaling with $\langle \tau \rangle$.
Instead, the observed diffusivity is excellently explained by our theory (solid line), Eq.~\eqref{eq:D_L}. The theory explains the Pe-dependent $D_L$s of active tracers of different sizes. We note that the tracer follows the Brownian-limit diffusivity $D_L^0$ (dashed line) only when the confinement effect overwhelms the hopping (filled symbols at Pe=0).

\section{Active diffusion for large particles}\label{sec5}

\begin{figure*}
\includegraphics[width=0.7\textwidth]{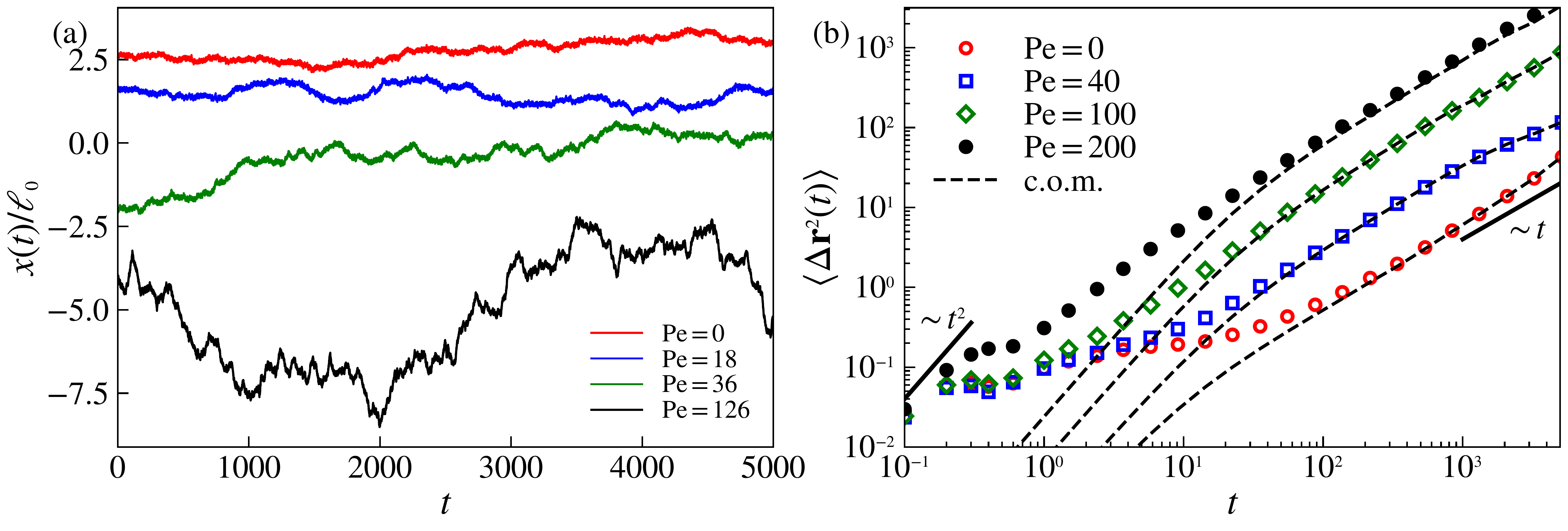}
\caption{\label{fig4}
(a) The sample trajectory and (b) MSD curves for large tracers of size $\sigma_{\mathrm{tr}} = 10$.
}
\end{figure*}

Finally, we consider the limiting case in which a tracer of size larger than the mean mesh size is confined to the network (Fig.~\ref{fig1}d). 
When $\sigma_\mathrm{tr} > \ell_0$, the tracer is too tightly squeezed within a polymer mesh and needs huge activation energy to be released. 
Figure~\ref{fig4}a shows the sample trajectories of AOUPs of size $\sigma_\mathrm{tr}=2\ell_0$. For Pe conditions investigated up to 250, all the tracers exhibit confined diffusion within the trapped site. Only the amplitude of the fluctuation gets larger with increasing Pe.   
In this limiting situation with no hopping transitions, the tracer dynamics essentially reflects the polymer network dynamics, driven by a large Pe. 

We plot the corresponding MSDs in Fig.~\ref{fig4}b. The dynamic response is summarized in the following. For $t\sim O(10^{-1})$ all tracers experience the underdamped ballistic motion regardless of Pe condition. After this timescale, the tracers  exhibit a typical confinement-induced dynamics for $t\sim 10^0$. Up to this regime the diffusion dynamics is similar to the underdamped Brownian particle confined to a strong harmonic trap. Then they have polymer-involved Pe-dependent anomalous diffusion in the next regime before entering the Fickian dynamics at later times. In this regime, tracers having zero or a small Pe capture the collective dynamics of the polymer network, which is sort of a negative viscoelastic feedback against the tracer's local fluctuating motion. A very similar dynamic response can be found for an AOUP cross-linker in a simpler polymer network system (see the simulation and analytic theory in Ref.~\cite{Joo2020ABP}). Meanwhile, if the tracers are fluctuating with a high Pe condition (e.g., Pe=100 \& 200 in the plot), the collective dynamics are masked because the AOUP-driven drift of the entire system becomes significant, and the corresponding regime is simply a cross-over toward the Fickian regime. Note that despite the confined motion, the particles illustrate Fickian dynamics at large times.   
For these strongly trapped active tracers, the Pe-dependent Fickian regime occurs because the total system has an accelerated drift by the active fluctuation. The agreement of the MSD with that of the center-of-mass (dashed line) supports this interpretation.  

%Fig.~\ref{fig::large_traj} shows the sample trajectories of large particles and Fig.~\ref{fig::large_MSD} shows its EA-TAMSD. For the large particle, the particle trajectories for each particles have the same profile because the fluctuation of the particle in each polymer network meshes are small compare to the fluctuation of polymer network itself. 

%%%%%%%%%%%%% Fig 4 %%%%%%%%%%%%%%%%%%%%%%%%%%%%%%%%%%%%%%%%%%
%%%%%%%%%%%%%%%%%%%%%%%%%%%%%%%%%%%%%%%%%%%%%%%%%%%%%%%%%%%%%%%%%%

\section{Discussion: Long-time diffusivity}\label{sec6}

%%%%%%%%%%%%% FIg 5 %%%%%%%%%%%%%%%%%%%
\begin{figure}
\includegraphics[width=8cm]{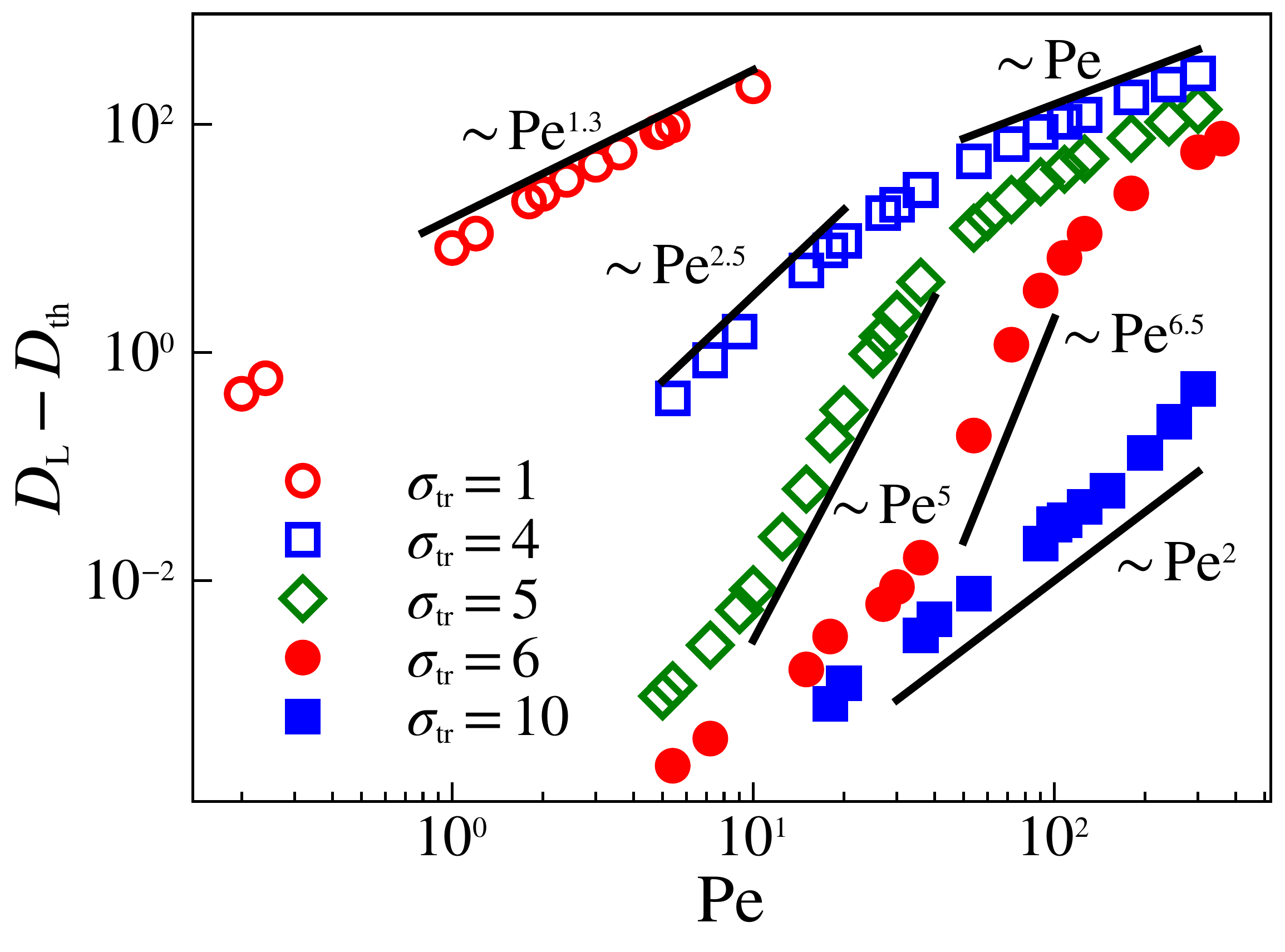}
\caption{\label{fig5}The long-time diffusivity $D_L$ as a function of P\'eclet number Pe. %The long-time diffusivity increases as activity increases, but it has at most linear order of activity at large scales.
}
\end{figure}
%%%%%%%%%%%%%%%%%%%%%%%%%%%%%%%%%%%%%%%%

The active diffusion of AOUPs confined to polymer networks turns out to have rich, distinctive dynamics, which depends on different tracer sizes and active forces.  
In Fig.~\ref{fig5}, a comprehensive picture of the observed results is presented in terms of our central quantity, the long-time diffusivity $D_L$, subtracted by the thermal diffusion that is independent of Pe. The active part of $D_L$ scales as $D_L-D_\mathrm{th}\propto \operatorname{Pe}^2$ with Pe for AOUPs in free space [Eq.~\eqref{eq8}].  

The long-time diffusivity, particularly for the mesh-sized AOUPs, increases at most with the linear order of $\operatorname{Pe}$ for large Pe values, below which more complex scaling behaviors are found.
This is an intriguing behavior. 
%The long-time diffusivity of AOUPs in free space has the second leading order term of activity, i.e., $D \sim \operatorname{Pe} ^2$ [see Eq.~\eqref{eq8}].
%However, our result reveals that the $D_L \sim \operatorname{Pe}^\nu $ with $\nu < 2$ for large $\operatorname{Pe}$. The Fig.~\ref{diffusivity_vp} shows that the exponent $\nu$ decreases as $\operatorname{Pe}$ increases. For large activity, the exponent $\nu < 1$ which means the diffusivity has at most linear order of activity.

We find that $D_L$ has distinct scaling relations with respect to Pe for three different regimes, which is summarized as
\begin{eqnarray}\label{eq:dis}
D_L \sim
\begin{cases}
C(\text{Pe}) \text{Pe}^2 &,~\sigma_\text{tr} \ll \ell_0 \\
\frac{\left \langle \ell^2 \right \rangle/6}{\langle \tau \rangle + \langle \tau_\mathrm{fl} \rangle} + D_\text{com} &,~\sigma_\text{tr} \approx \ell_0\\
%\frac{\left \langle \ell^2 \right \rangle}{\langle \tau_\mathrm{fl} \rangle} + D_\text{com} &,~\sigma_\text{tr} \approx \ell_0, \text{Pe} \gg 1 \\
\text{Pe}^2 &,~\sigma_\text{tr} \gg \ell_0.
\end{cases}
\end{eqnarray}
For small AOUPs, the long-time motion is essentially the free diffusion, suppressed by the factor $C(\text{Pe})$ that effectively incorporates the obstacle contributions.
We find the prefactor well matched with $C\sim \text{Pe}^{-0.7}$ (see Fig.~A1 in the Appendix~\ref{appedixa}) yields the theory in Eq.~\eqref{eq:dis}, $D_L \sim C(\text{Pe}) \text{Pe}^2 \sim \text{Pe}^{1.3}$ in good agreement with the simulation result, as shown in Fig.~\ref{fig5}.
For very large AOUPs, the long-time diffusion originates from the center-of-mass motion of the entire system, thereby simply following the scaling $D_\text{com}\sim \text{Pe}^2$ (see the filled square symbol in Fig.~\ref{fig5} and the Appendix~\ref{appendixe} (Fig.~A4d)).
Very intriguing features are found in the mesh-sized AOUPs, the diffusivity of which is captured by Eq.~\eqref{eq:D_L} [or Eq.~\eqref{eq:dis}], where the exponent $\nu$ of $D_L\sim \text{Pe}^\nu$ abruptly changes (decreases) at around $\text{Pe}^\ast\approx50$ for tracers of $\sigma_\text{tr}=4$ and 5. The latter ($\text{Pe}^\ast$) corresponds to the threshold active force beyond which trapped tracers are released and undergo the hopping-based diffusion in a narrow channel. We notice that this rich and complex scaling behavior of $D_L$ mainly depends on the trapped time $\langle \tau \rangle$ for $\text{Pe} < \text{Pe}^\ast$ (because $\langle \tau \rangle\gg \langle \tau_\mathrm{fl}\rangle$) and $\langle l^2\rangle$ (see Eq.~\eqref{eq:dis}). If $\text{Pe} > \text{Pe}^\ast$, the $D_L$'s scaling relation nontrivially depends on the combination of $\langle l^2\rangle$, $\langle \tau\rangle$, $\langle \tau_\text{fl} \rangle$, and $D_\mathrm{com}$.  
For these four quantities, we plot each term's Pe-dependency in the Appendices~\ref{appendixd} and \ref{appendixe}.
%the Pe-dependecy of $\langle \ell^2 \rangle$, $\langle \tau \rangle$, $\langle \tau_\text{fl} \rangle$, and $D_\mathrm{com}$. 

We also remark that, for the mesh-sized tracers, the exponent $\nu$ tends to be maximal in the range $\text{Pe}^\dagger < \text{Pe} < \text{Pe}^\ast$, where $\text{Pe}^\dagger$ denotes the activation force that starts to trigger a fluctuation of trapped tracers in conjugation with polymer networks.
For example, for the tracers of size $\sigma_\text{tr}=6$, we find $D_L \sim \text{Pe}^{6.5}$ for $\text{Pe}^\dagger = 50 < \text{Pe} < \text{Pe}^\ast = 100$, otherwise $D_L \sim \text{Pe}^2$. This allows us to interpret three different regimes of the long-time diffusion in terms of Pe: (i) For $\text{Pe} < \text{Pe}^\dagger$, tracers are mostly trapped and the dynamics is predominantly determined by the network center-of-mass motion, $D_L \sim \text{Pe}^2$. (ii) For $\text{Pe}^\dagger < \text{Pe} < \text{Pe}^\ast$, tracers are still trapped but more activated, resulting in the dynamics affected by a large fluctuation of the trapped tracers and responsive polymer networks, thereby having a large $\nu$. (iii) For $\text{Pe} > \text{Pe}^\ast$, highly activated tracers are released from confining meshes and undergo a normal diffusion, $D_L \sim \text{Pe}^2$. 
Therefore, we attribute the nontrivial scaling behavior of $D_L \sim \text{Pe}^\nu$ with $\nu>2$ to the fluctuations of the activated tracers as well as the responsive polymer network. 

\section{Conclusions}\label{sec7}

We have investigated the transport dynamics of active tracers in polymer networks for varying active forces and mesh-to-particle sizes. 
For small tracers, the long-time diffusive motion was found to be similar to normal active motion with a reduced diffusivity, due to the polymer network as obstacles that make the effective free space smaller. For mesh-sized tracers, the dynamics is the result of competition between active persistent, superdiffusive motion and confined motion by polymer network meshes. The activity helps the particle escape from polymer network meshes and run longer jump lengths. The multiple jump events are the characteristic of these intermediate-sized active particles. The long-time behavior of trapped and hopping motion in the polymer network was explained using the random walk analogy. The mean sojourn (trapped) time, flight time and mean square jump lengths turned out to be important quantities that successfully explain the long-time diffusivities.

We also discussed the behavior of long-time diffusivity $D_L$ of the mesh-sized tracer particles depending on the P\'eclet number, which is quite new. Our comprehensive study revealed that there is an intriguing range of Pe that is intertwined with different physics. We found two characteristic P\'eclet numbers, namely $\text{Pe}^\dagger$ and $\text{Pe}^\ast$, at which the scaling behavior of $D_L$ changes. 
For a small $\text{Pe} < \text{Pe}^\dagger$, the center-of-mass motion of the polymer network prevails where the tracer is trapped in the network mesh, thereby exhibiting $D_L \sim \text{Pe}^2$.     
For $\text{Pe}^\dagger < \text{Pe} < \text{Pe}^\ast$, the highly activated but still trapped tracer fluctuates largely including a viscoelastic feedback response from the network, which leads to $D_L \sim \text{Pe}^\nu$ with $\nu>2$. For a large $\text{Pe} > \text{Pe}^\ast$, the tracer is released from the confinement and undergoes a hopping diffusion, and $D_L$ shows a complicated Pe-dependency such that $D_L \sim \text{Pe}^\nu$ with  $\nu(\operatorname{Pe})\gtrsim1$ depending on Pe.
%, $D_L \sim \text{Pe}^2$.
This nontrivial and distinct scaling behavior of $D_L(\text{Pe})$ is revealed markedly only for tracers of size comparable with the network mesh size, which can be utilized in practice to examine a pore size or mesh size of gels with active tracers of known size.

Transport of active tracers in a polymer network is a paramount subject that requires a better understanding. 
Many studies of various active particles in free spaces are carried out, and some studies considered active tracers in polymer solutions~\cite{rajarshi2017Janustracer,yuan2019PCCP,du2019softmatter}. Although the environments are viscoelastic polymers, the networked system and the solution system reveal different results. For large particles, a polymer solution acts as viscous fluid but a polymer network confines the particle. 
Our results incorporating this network confinement into the active diffusion revealed intriguing mechanism, which opens a new avenue to better understand the active transport in omnipresent gels.  

\begin{acknowledgments}
This work was supported by the National Research Foundation (NRF) of Korea (No.~2020R1A2C4002490).
W.K.K. acknowledges the financial support from the KIAS Individual Grants (CG076002) at Korea Institute for Advanced Study.
We acknowledge the Center for Advanced Computation at Korea Institute for Advanced Study for providing computing resources for this work.
\end{acknowledgments}

\appendix
    
%\section{APPENDIX}
\renewcommand{\thefigure}{A\arabic{figure}}
\setcounter{figure}{0}

\section{The Pe-dependence of $C$}\label{appedixa}
\begin{figure}%[b]
    \includegraphics[width=6cm]{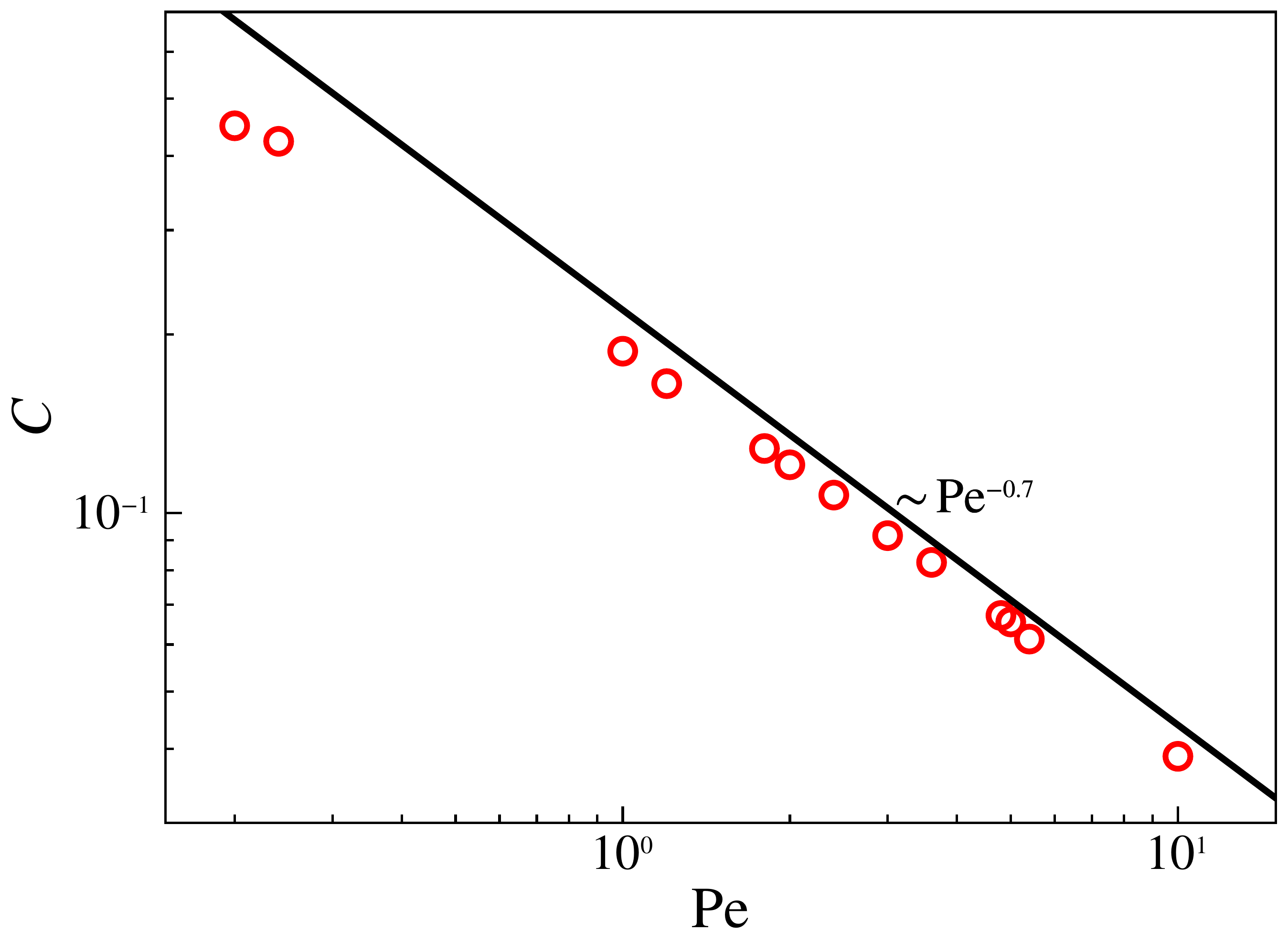}
    \caption{\label{Pe_C_data} The fitting parameter for the prefactor $C$ as a function of Pe. We show $C\sim \text{Pe}^{-0.7}$ as a guide line.}
\end{figure}

%In Fig.~\ref{fig2}b, we have fitted the MSDs for small active tracers with the free-space formula given in Eq.~\eqref{msdfree} multiplied by the fit parameter $C(\operatorname{Pe})$. 
Figure~\ref{Pe_C_data} shows $C$ vs. Pe, which was discussed in Fig.~\ref{fig2}b. The prefactor $C$ shows a power-law dependence on Pe as guided by $C \sim \text{Pe}^{-0.7}$. 
%%%%%%%%%%% Fig Appendix 1 %%%%%%%%%%%%%%
%%%%%%%%%%%%%%%%%%%%%%%%%%%%%%%%%%%%%%%%

\section{Trapped-and-hopping}\label{appendixb}
The trapped-and-hopping dynamics features the transport dynamics of mesh-sized tracers. However, an appropriate and accurate definition for a trapped state and a hopping state is not well established and specifically depends on systems. 
We here adopted the skeletonized trajectory of the tracer and detected the hopping events as follows.

First, we considered a tracer's trajectory $x_j(t)$ for time $t = 0, t_0, 2t_0, 3t_0 \cdots $.
Here we use $j=1,2,3$ where $x_1, x_2, x_3$ denote $x,y,z$-component trajectories. 
For each $x_j (t)$, we skeletonized the trajectory using an edge-preserving filter called the bilateral filter~\cite{banterle2012low,tomasi1998bilateral}, 
\begin{align}
\tilde{x}_j(t_k) =  {\frac {1}{W_{p}}}\sum _{t_i \in \Omega } x_{j}(t_i) f\qty(\qty| x_{j}(t_i) - x_j(t_k)|) g(|t_{i}-t_k|),
\end{align}
where $ W_{p}=\sum _{t_i \in \Omega } f\qty(\qty| x_{j}(t_i) - x_j(t_k)|)$ is the normalization factor, $f$ and $g$ are the range and domain kernel for smoothing, respectively. 
Our bilateral filter is based on the gaussian weight, $f(x) = e^{-x^2/\sigma_x ^2}$ and $g(t) = e^{-t^2/\sigma_t ^2}$, with $\sigma_x =5$ and $\sigma_t=1$. 
%We skeletonized the trajectories by using this bilateral filter twice.
%So, our skeletonized trajectory is $\tilde{\tilde{x}}_j (t)$. For simplicity, we will just write tilde symbol only once. So from now on, the skeletonized trajectory $\tilde{x}$ states for $\tilde{\tilde{x}}$.

\begin{figure}%[b]
    \includegraphics[width=6cm]{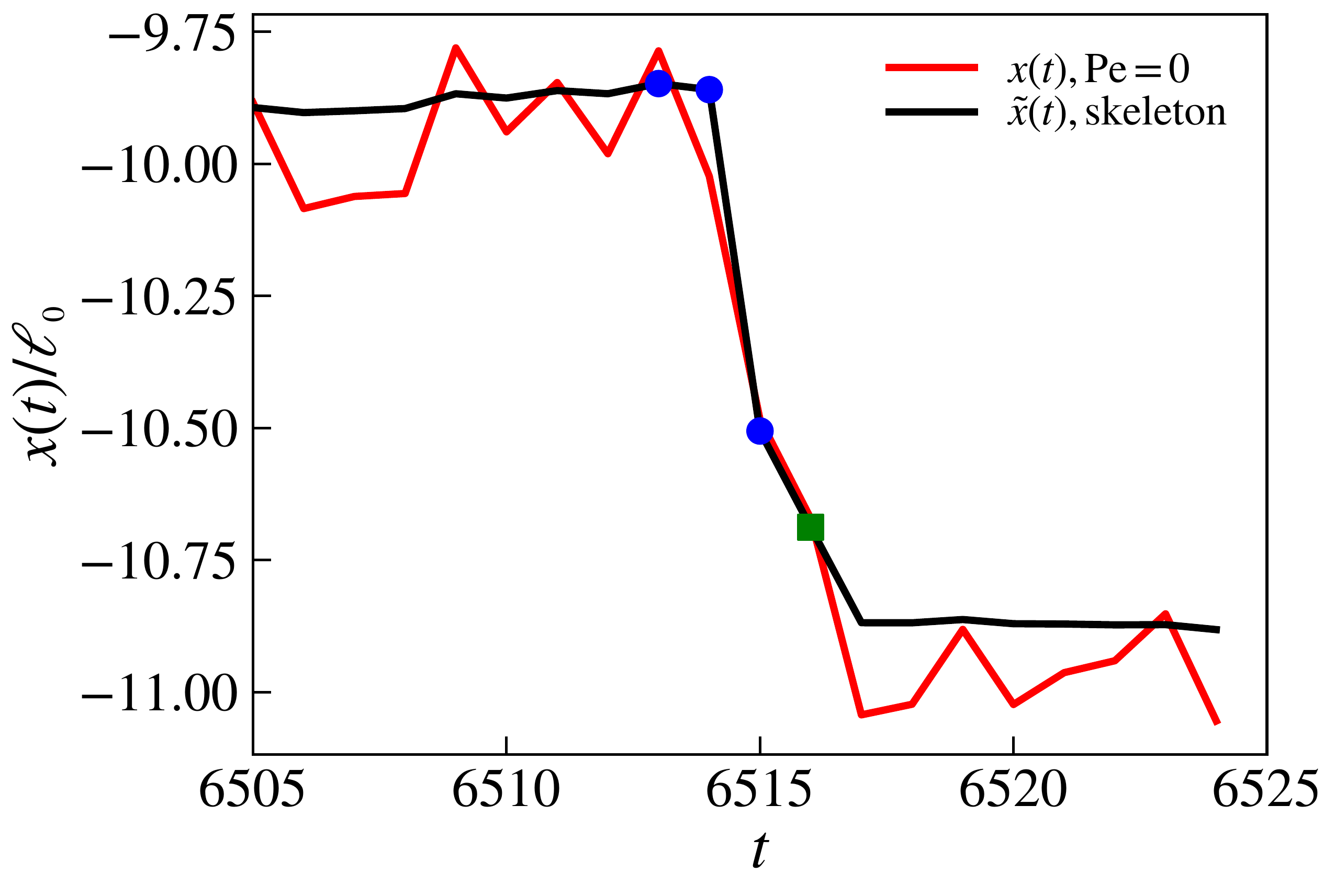}
    \caption{\label{hopping_fig}The original and skeletonized trajectory of the tracer. The blue circles are the dots that satisfies $\qty| \tilde{x}(t_i + 2t_0) - \tilde{x}(t_i - t_0) | > 3\sigma_0 $ and green square dots are additional hopping points with the algorithms in the text.}
\end{figure}

Using the skeletonized trajectory $\tilde{x}(t_i)$, we then found the hopping spots. 
When the skeletonized trajectory goes to a directional motion for a while, where the distance within the directional motion should be at least comparable to the mesh size and be within a certain time, we regard that the hopping events occur. The time should not be very large to rule out the thermal fluctuation of the polymer network, but it should not be very small not to miss some hopping points. For this, we set a criterion that the distance is more than $60\%$ of the mesh size, $3\sigma_0$ within time $3t_0$.
In other words, if $\qty| \tilde{x}_j(t_i + 2t_0) - \tilde{x}_j(t_i - t_0) | > 3\sigma_0$, then $t_i$ represents the time in a hopping event. 

Next, we determined the beginning and end time for hopping. Since our definition of hopping is the directional motion for short times, for each hopping time we check if the direction of the motion changes or not in time. If the time $t_i$ is in a hopping event and the direction (or sign) of $x_j(t_i)-x_j(t_{i-1})$, $\tilde{x}_j(t_i)-\tilde{x}_j(t_{i-1})$, $x_j(t_{i+1})-x_j(t_{i})$ and  $\tilde{x}_j(t_{i+1})-\tilde{x}_j(t_{i})$ are the same, then $t_{i}$ is also in a hopping event. We iteratively imposed this condition to find the end time of the hopping event. Likewise, to find the start time for hopping, we regarded that $t_{i-1}$ is in a hopping event if $t_{i}$ is in the hopping event and $x_j(t_i)-x_j(t_{i-1})$, $\tilde{x}_j(t_i)-\tilde{x}_j(t_{i-1})$, $x_j(t_{i+1})-x_j(t_{i})$ and $\tilde{x}_j(t_{i+1})-\tilde{x}_j(t_{i})$ have the same direction.

Since the trajectory is in 3D, we defined that $t_k$ is in a trapped state if $t_k$ is not in a hopping state for any $x_j$. Then one can find the sequence $t_k, t_{k+1}, \cdots, t_{m}$ such that $t_{k-1}$ and $t_{m+1}$ are not in the trapped state, but $t_{k}, \cdots , t_m$ is in the trapped state. 
Therefore, we defined the trapped time as $t_{m+1} - t_k$.

For the flight time in 3D, when we detected the sequence of time, $t_k, t_{k+1}, \cdots, t_{m}$ such that $t_{k-1}$ and $t_{m+1}$ are in a trapped state, but $t_{k}, \cdots , t_m$ is not in the trapped state, we defined the flight time as $\tau_{\mathrm{fl}}=t_{m+1}-t_{k}$ and flight length as $\sqrt{\sum_{i}\qty( \tilde{x}_i(t_{m+1})-\tilde{x}_i(t_{k}) )^2 }$. 

%For the passive or low $\operatorname{Pe}$ case, the hopping toward certain direction right after the hopping toward the another direction is rare event, so we can ignore this possibility. 
When $\operatorname{Pe}$ is very large, a hopping event can be a sequence of smaller mesh-to-mesh hoppings as the combination of $x_1$, $x_2$, and $x_3$ directions. In this case, we detected a successive sequence of hopping events. For example, when the time $t_{k}, t_{k+1}, \cdots, t_{m}$ is the hoppping time for $x_{1}$ direction and $t_{m+1}, t_{m+2}, \cdots, t_{q}$ is hopping time for $x_{2}$ direction, while $t_{m-1}$ and $t_{q+1}$ is not in a hopping state, we regarded $t_{k}, t_{k+1}, \cdots, t_{q}$ as one hopping event.

%%%%%%%%%%%%%%%%%%% FIG A3 %%%%%%%%%%%%%%%%%%%
\begin{figure}\label{figa3}
\includegraphics[width=6cm]{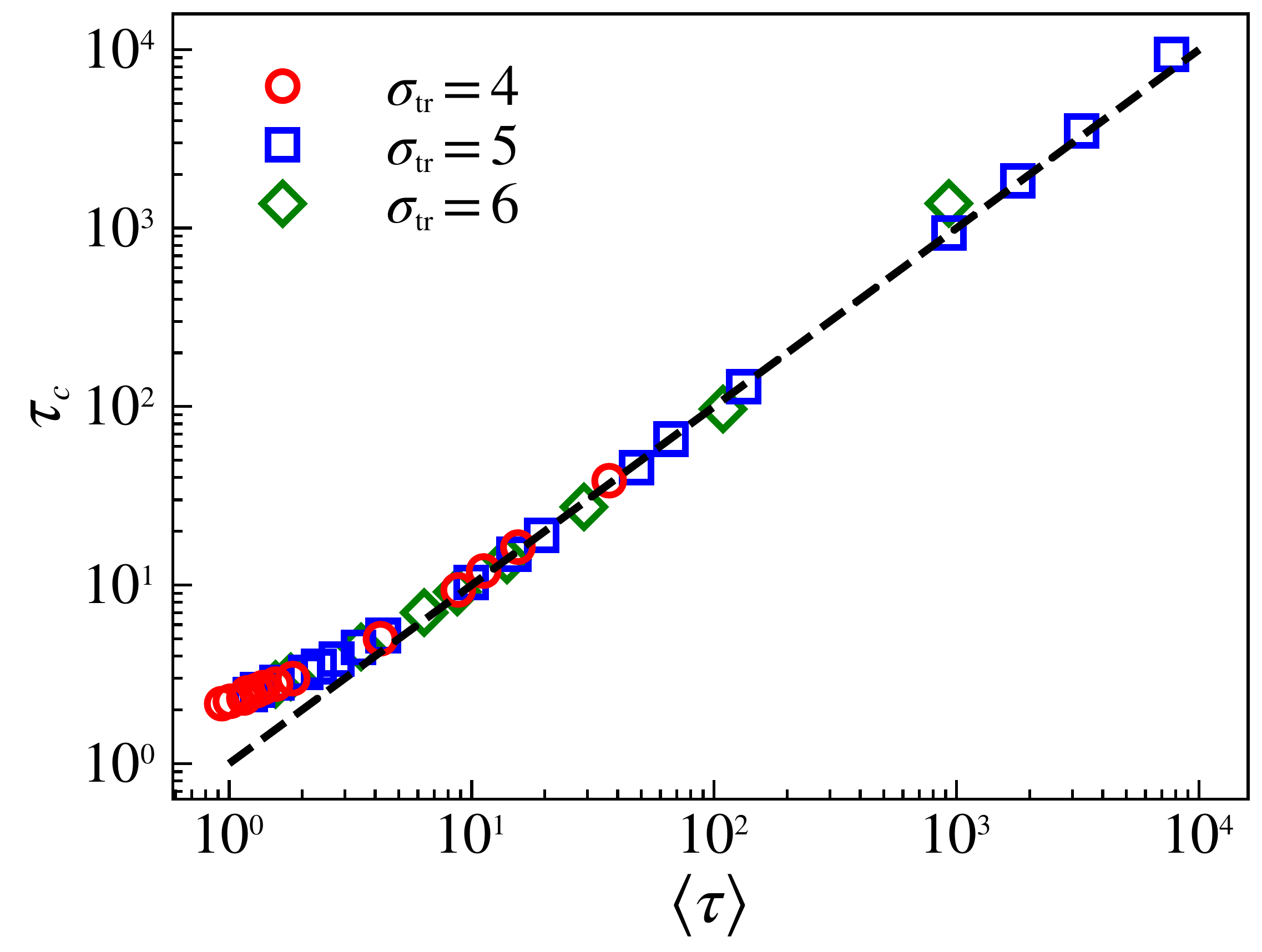}
\caption{\label{mtt_vs_fit} The mean trapped time and the corresponding fit parameter. }
\end{figure}
%%%%%%%%%%%%%%%%%%%%%%%%%%%%%%%%%%%%%%%%%%%%

\section{Mean trapped time and fit parameter}\label{appendixc}

In Fig.~A3, we show the mean trapped time $\langle \tau \rangle$ and the fitted trapped time $\tau_c$, which was discussed in Fig.~3b--c.

%%%%%%%%%%%%%%%%%%%%%%%%%% FIG A4 %%%%%%%%%%%%%%%%%%%%%%
\begin{figure*}
\includegraphics[width=0.7\textwidth]{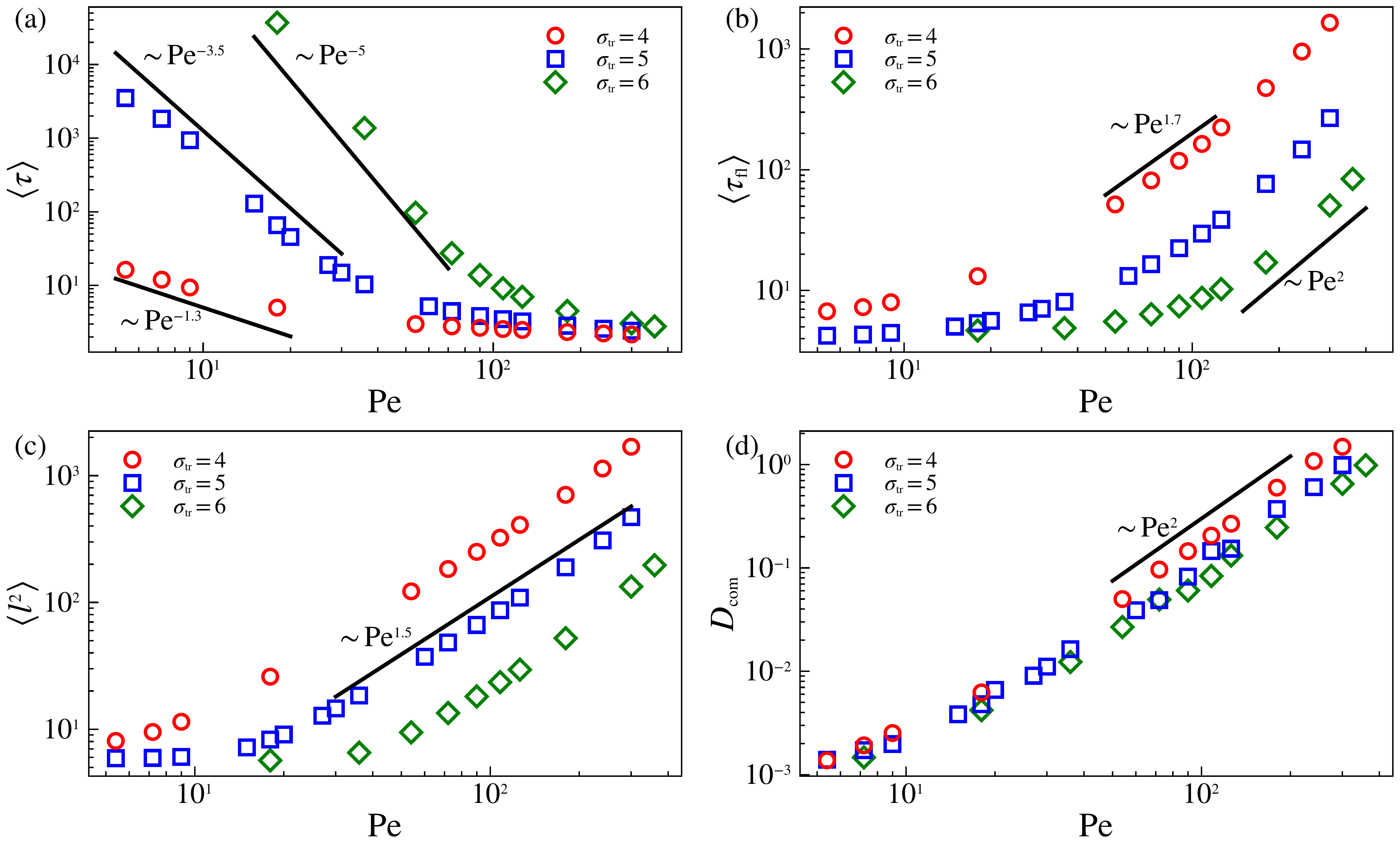}
\caption{(a) The mean trapped time as a function of Pe. 
(b) The mean flight time as a function of Pe.
(c) The mean-squared flight length as a function of Pe.
(d) The long-time diffusivity of the system center-of-mass $D_\text{com}$ as a function of Pe.
\label{Pestatistics}
}
\end{figure*}
%%%%%%%%%%%%%%%%%%%%%%%%%%%%%%%%%%%%%%%%%%%%%%%

\section{Mean trapped time, mean flight length and time}\label{appendixd}

In Figs.~\ref{Pestatistics}a--c, we show the mean trapped time $\langle \tau \rangle$, the mean flight time $\langle \tau_\text{fl} \rangle$, and the mean-squared flight length $\langle \ell^2 \rangle$ as function of Pe, which was discussed in Sec. VI.

\section{Long-time diffusivity of the entire system}\label{appendixe}

In Fig.~\ref{Pestatistics}d, we show the long-time diffusivity of the system center of mass $D_\text{com}$ as a function of P\'eclet number Pe, which follows the scaling $D_\text{com} \sim \text{Pe}^2$.

\iffalse
\begin{figure}
    \includegraphics[width=6cm]{Pel2_log_log_stix.pdf}
    \caption{The mean-squared flight length as a function of Pe.}
\end{figure}

\begin{figure}
    \includegraphics[width=6cm]{Petaufl_log_log_stix.pdf}
    \caption{The mean flight time as a function of Pe.}
\end{figure}

\begin{figure}
\includegraphics[width=7cm]{comDL_PElog_stix_all.pdf}
    \caption{The long-time diffusivity of the system center of mass $D_\text{com}L$ as a function of P\'eclet number Pe.}
\end{figure}
\fi

\newpage %Just because of unusual number of tables stacked at end
\bibliography{ref}% Produces the bibliography via BibTeX.

\end{document}